\documentclass[debug,overfull,figures]{epl}

\title{Quasiparticles and quantum phase transition in universal
low-temperature properties of heavy-fermion metals}
\shorttitle{Universal low-temperature properties}

\author{V. R. Shaginyan\inst{1,2}\thanks {Email:
\email{vrshag@thd.pnpi.spb.ru}} \and A. Z. Msezane\inst{2} \and V.
A. Stephanovich\inst{3}\thanks{Email:
\email{stef@math.uni.opole.pl} and Homepage:
http://cs.uni.opole.pl/$\sim$stef} \and E. V. Kirichenko\inst{3}}
\shortauthor{V.R.Shaginyan \etal}

\institute{ \inst{1} Petersburg Nuclear
Physics Institute, Gatchina, 188300, Russia\\
\inst{2} CTSPS, Clark Atlanta University, Atlanta, Georgia 30314,
USA \\
\inst{3} Opole University, Institute of Mathematics and Informatics,
Opole, 45-052, Poland}

\begin{document}
\maketitle
\begin{abstract}
We demonstrate, that the main universal features of the low
temperature experimental $H-T$ phase diagram of CeCoIn$_5$ and
other heavy-fermion metals can be well explained using Landau
paradigm of quasiparticles. The main point of our theory is that
above quasiparticles form so-called fermion-condensate state,
achieved by a fermion condensation quantum phase transition
(FCQPT). When a heavy fermion liquid undergoes FCQPT, the
fluctuations accompanying above quantum critical point are strongly
suppressed and cannot destroy the quasiparticles. The comparison of
our theoretical results with experimental data on CeCoIn$_5$ have
shown that the electronic system of above substance provides a
unique opportunity to study the relationship between quasiparticles
properties and non-Fermi liquid behavior.
\end{abstract}

Although much theoretical efforts have been devoted to understand
the non-Fermi liquid (NFL) behavior of heavy fermion (HF) metals
using the concept of quantum critical points, the problem is still
far from its complete understanding since the experimental systems
display serious discrepancies with the theoretical predictions
\cite{stew}. Common belief is that a quantum critical point (QCP)
is the point where a second order phase transition occurs at
temperature $T\to 0$, and where both thermal and quantum
fluctuations are present destroying quasiparticles and generating a
new regime around the point of instability between two stable
phases \cite{col}. Recent experimental studies of the CeCoIn$_5$ HF
metal provide valuable information about the NFL behavior near
possible QCP due to its excellent tunability by a pressure $P$
and/or a magnetic field $H$ \cite{pag,pag2,bi,ronn}. The
experimental studies have shown that besides a complicated $H-T$
phase diagram, the normal and superconducting properties around the
QCP exhibit various anomalies. One of them is power (in both $T$
and $H$) variation of the resistivity and heat transport
\cite{pag,pag2,bi,ronn,mal,bauer}, inherent to  both NFL and Landau
Fermi liquid (LFL) regimes. The other one is a continuous magnetic
field evolution of a superconductive phase transition from the
second order to the first one \cite{izawa,bian}. Above anomalous
power laws can be hardly accounted for within scenarios based on
the QCP occurrence with quantum and thermal fluctuations. For
example, the divergence of the normal-state thermal expansion
coefficient, $\alpha/T$ is stronger than that in the 3D itinerant
spin-density-wave (SDW) theory, but weaker than that in the 2D SDW
picture \cite{pag1}. This brings the question of whether the
fluctuations are responsible for the observed behavior, and if they
are not, what kind of physics determines the above anomalies?
Fortunately, the direct observations of quasiparticles in
CeCoIn$_5$, CeIrIn$_5$, and Sr$_3$Ru$_2$O$_7$ have been reported
recently \cite{pag2,fujim,ronn1}. However, if the quasiparticles do
exist, why they are not suppressed by the fluctuations? Moreover,
these recent facts contradict strongly to the theoretical
investigations, where quantum phase transitions responsible for the
NFL behavior considered to be of a second kind so that the
quasiparticles are inhibited near these phase transitions
\cite{stew,col}.

In this letter we show that these problems can be resolved within
Landau quasiparticle picture providing that quasiparticles form the
so-called fermion-condensate (FC) state \cite{ks} emerging behind
the fermion condensation quantum phase transition (FCQPT)
\cite{ams}. We show that near FCQPT the fluctuations are strongly
suppressed while quasiparticles are "protected" from above
fluctuations by the first order phase transition. We analyze the
experimental $H-T$ phase diagram of CeCoIn$_5$ and show that its
main universal features can be well understood within the theory
based on FCQPT. We demonstrate that the electronic system of
CeCoIn$_5$ can be shifted from the ordered to disordered side of
FCQPT by a magnetic field; therefore giving a unique possibility to
study the relationship between quasiparticles and NFL behavior.

To study the low temperature universal features of HF metals, we
use the notion of HF liquid in order to avoid the complications
associated with the crystalline anisotropy of solids. This is
possible since we consider the (universal) behavior related to the
power-law divergences of observable variables like the effective
mass, thermal expansion coefficient etc. These divergences are
determined by small (as compared to those from unit cell of a
corresponding reciprocal lattice) momenta transfer so that the
contribution from larger momenta can be safely ignored. Let us
consider HF liquid characterized by the effective mass $M^*$. Upon
applying the well-known equation, we can relate $M^*$ to the bare
electron mass $M$ \cite{lif_pit,pfit}
$M^*=M/(1-N_0F^1(p_F,p_F)/3).$ Here $N_0$ is the density of states
of a free electron gas, $p_F$ is Fermi momentum, and $F^1(p_F,p_F)$
is the $p$-wave component of Landau interaction amplitude. Since
LFL theory implies the number density in the form $x=p_F^3/3\pi^2$,
we can rewrite the amplitude as $F^1(p_F,p_F)=F^1(x)$. When at some
$x=x_{\rm FC}$, $F^1(x)$ achieves some critical value, the
denominator tends to zero so that the effective mass diverges at
$T=0$. Beyond the critical point $x_{\rm FC}$ the denominator
becomes negative making the effective mass negative. To avoid
physically meaningless states with $M^*<0$, the system undergoes
FCQPT with FC formation in the critical point $x=x_{\rm FC}$.
Therefore, behind the critical point $x_{\rm FC}$ the quasiparticle
spectrum is flat, $\varepsilon({\bf p})=\mu$, in some region
$p_i\leq p\leq p_f$ of momenta, while the corresponding occupation
number $n_0({\bf p})$ varies continuously from 1 to 0, $0<n_0({\bf
p})<1$ \cite{ks}. Here $\mu$ is a chemical potential. To
investigate the FC state at $T=0$, we apply weak BCS-like
interaction with the coupling constant $g$ and see what happens
with both the superconducting gap $\Delta$ and the superconducting
order parameter $\kappa({\bf p})$ as $g \to 0$. Let us write the
usual pair of equations for the Green's functions $F^+({\bf
p},\omega)$ and $G({\bf p},\omega)$ (see e.g. ref. \cite{lif_pit})
\begin{equation}\label{zui2}
F^+=\frac{-g\Xi^*}{(\omega -E({\bf p})+i0)(\omega +E({\bf p})-i0)};
\,\,G=\frac{u^2({\bf p})}{\omega -E({\bf p})+i0}+\frac{v^2({\bf
p})}{\omega +E({\bf p})-i0},
\end{equation}
where $E^2({\bf p})=\xi^2({\bf p})+\Delta^2$, $\xi({\bf
p})=\varepsilon({\bf p})-\mu$,  and the superconducting gap,
\begin{equation}\label{zui3}
\Delta=g|\Xi|,\quad i\Xi= \int\int_{-\infty }^{\infty }F^+({\bf
p},\omega)\frac{d\omega d{\bf p} }{(2\pi)^4}.
\end{equation}
Here $v^2({\bf p})=(1-{\xi({\bf p})}/{E({\bf p})})/2,\, v^2({\bf
p})+u^2({\bf p})=1$, and simple transformations give
\begin{equation}\label{zui6} \xi({\bf p})=\Delta\frac{1-2v^2({\bf p})}{2\kappa ({\bf p})};\,\,\,
\frac{\Delta}{E({\bf p})}=2\kappa({\bf p}),
\end{equation}
with $\kappa ({\bf p})=u({\bf p})v({\bf p})$. Next we observe from
eqs. (\ref{zui3}) and (\ref{zui6}) that
\begin{equation}\label{zui7}
i\Xi=\int_{-\infty }^{\infty }F_0^+({\bf p},\omega )\frac{d\omega
d{\bf p}}{(2\pi)^4}=i\int\kappa({\bf p})\frac{d{\bf p}}{(2\pi)^3}.
\end{equation}
It follows from eqs. (\ref{zui3}), (\ref{zui6}) and (\ref{zui7})
that when $g\to0$ the superconducting gap $\Delta\to0$, while
$\xi=0$ and the dispersion $\varepsilon ({\bf p})$ becomes flat,
providing that $\kappa({\bf p})$ is finite in some region $p_i\leq
p\leq p_f$, making $\Xi$ finite. Thus, in the state with FC $\Delta$
can vanish while parameters $\kappa({\bf p})$ and $\Xi$ are finite.
Taking into account eqs. (\ref{zui3}) and (\ref{zui6}) we represent
eqs. (\ref{zui2}) as follows
\begin{equation}\label{zui8}
F^+=-\frac{\kappa({\bf p})}{\omega -E({\bf
p})+i0}+\frac{\kappa({\bf p})}{\omega +E({\bf p})-i0};\,\,\,
G=\frac{u^2({\bf p})}{\omega -E({\bf p})+i0}+\frac{v^2({\bf
p})}{\omega +E({\bf p})-i0}.
\end{equation}
It is directly seen from eqs. (\ref{zui8}) that in the FC state at
$g \to 0$, the equations for functions $F^+({\bf p},\omega )$ and
$G({\bf p},\omega )$ take the following form in the region where
$\kappa({\bf p})\neq 0$
\begin{equation}\label{zui9}
F^+({\bf p},\omega )=-\kappa({\bf p})\left[ \frac{1}{\omega +i0}
-\frac{1}{\omega -i0}\right];\,\,\, G({\bf p},\omega
)=\frac{u^2({\bf p})}{\omega +i0}+\frac{v^2({\bf p})}{\omega -i0}.
\end{equation}
Here, the factors $v^2({\bf p})$, $u^2({\bf p})=1-v^2({\bf p})$ are
determined by the condition $\varepsilon({\bf p})=\mu$ when $p_i\leq
p\leq p_f$. Upon integrating $G({\bf p},\omega )$ over $\omega$ we
obtain that $v^2({\bf p})=n({\bf p})$, where $n({\bf p})$ is the
quasiparticles distribution function. Taking into account the
well-known Landau equation $\delta E[n({\bf p})]/\delta n({\bf
p})=\varepsilon({\bf p})$, we observe that the equation determining
$n({\bf p})$ takes the form \cite{ks}
\begin{equation}\label{zui10}
\frac{\delta E[n({\bf p})]}{\delta n({\bf p})}=\mu;\quad  p_i\leq p
\leq p_f,
\end{equation}
where $E[n({\bf p})]$ is Landau functional \cite{lif_pit}. Equation
(\ref{zui10}) describes the state with FC characterized by the
superconducting order parameter $\kappa _0({\bf p})=\sqrt{n_0({\bf
p})(1-n_0({\bf p})}$ where the functions $n_0({\bf p})$ are
solutions of eq. (\ref{zui10}). It is instructive to construct
$F^+({\bf p},\omega)$ and $G({\bf p},\omega)$ when $g$ is finite
but small so that the functions $v^2({\bf p})$ and $\kappa({\bf
p})$ can be approximated by the solutions of eq. (\ref{zui10}). In
that case, $\Xi$, $\Delta$ and $E({\bf p})$ are given by eqs.
(\ref{zui7}), (\ref{zui3}) and (\ref{zui6}) respectively. Inserting
these into eqs. (\ref{zui8}) we obtain functions $F^+({\bf
p},\omega)$ and $G({\bf p},\omega)$. It is seen from eq.
(\ref{zui3}) that $\Delta$ is a linear function of the coupling
constant $g$. Since the transition temperature $T_c\sim \Delta$
tends to zero along with $g \to 0$, the order parameter
$\kappa({\bf p})$ of the FC state vanishes at any finite
temperature so that at $T>0$ the quasiparticle occupation number is
given by the Fermi-Dirac function which we represent in the form
$\varepsilon({\bf p},T) -\mu(T)=T\ln\{({1-n_0({\bf
p},T)})/({n_0({\bf p},T)})\}.$ Observing that at $T\to 0$ the
distribution function satisfies the inequality $0<n_0({\bf p})<1$
at $p_i\leq p\leq p_f$, we conclude that the logarithm is finite
(therefore $T\ln(...)\to 0$) and again arrive at eq. (\ref{zui10})
determining $n_0({\bf p})$. The entropy $S[n({\bf p},T)]$ is given
by the familiar expression \cite{lif_pit}
\begin{equation}
S[n({\bf p},T)]=-2\int[n({\bf p},T)\ln n({\bf p},T)+(1-n({\bf
p},T))\ln(1-n({\bf p},T))]\frac{d{\bf p}}{(2\pi)^3}.\label{ui12}
\end{equation}
It follows from eq. (\ref{ui12}) that the entropy $S_{\rm NFL}$
related to the special solution $n_0({\bf p})$ contains the
temperature independent term $S_0=S_{\rm NFL}(T \to 0)\sim
x(p_f-p_i)/p_F$. Thus, the function $n_0({\bf p})$ represents the
special solutions of both BCS and LFL equations determining the NFL
behavior of HF liquid with FC. Namely, contrary to conventional BCS
case, the FC solutions are characterized by infinitesimal value of
superconducting gap, $\Delta \to 0$, while both $\kappa({\bf p})$
and $\Xi$ remain finite and $S=0$. In contrast to the standard
solutions of the LFL theory, the special ones $n_0({\bf p})$ are
characterized by the entropy $S$ containing the temperature
independent term $S_0$. At $T \to 0$ both the normal state of the
HF liquid with the finite entropy $S_{0}$ and the BCS state with
$S=0$ coexist being separated by the first order phase transition,
where the entropy undergoes a finite jump $\delta S=S_0$. Due to
the thermodynamic inequality, $\delta Q\leq T\delta S,$ the heat
$\delta Q$ of the transition is equal to zero making the other
thermodynamic functions continuous. Thus, both at the FCQPT point
and behind it there are no critical fluctuations accompanying
second order phase transitions and suppressing the quasiparticles.
As a result, the quasiparticles survive and define the
thermodynamic properties of the HF liquid.

\begin{figure}
\onefigure[scale=0.25]{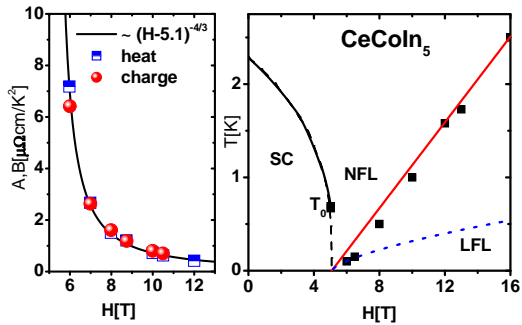} \caption{$H-T$ phase diagram of
CeCoIn$_5$. Right panel: Superconducting-normal phase boundary
\cite{bian} is shown by the solid and dashed lines with the solid
square showing the point where the superconducting phase
transition changes from the second to the first order. The dotted
line is given by eq. (\ref{tr1}) and represents the transition
$T^*(H)$ from the Landau Fermi liquid behavior (LFL) with the
$T^2$ regime in $\rho(T)$ to the $T^{2/3}$ one. The solid line
given by eq. (\ref{tr2}) represents the crossover $T^*(H)$ from
the $T^{2/3}$ regime in $\rho(T)$ \cite{pag} to the non-Fermi
liquid (NFL) behavior with $\rho(T)\propto T$. Experimental facts
obtained from resistivity measurements are shown by the solid
squares \cite{pag,pag2}. The left panel shows the magnetic field
dependence of $T^2$ Landau Fermi liquid coefficients of charge
$A(H)\propto (H-H_{c0})^{-4/3}$ and heat $B(H)\propto
(H-H_{c0})^{-4/3}$ transport with experimental data taken from
ref. \cite{pag,pag2}.} \label{fig1}
\end{figure}

On the basis of the above special solutions related to FC, we can
explain the main universal properties of the $H-T$ phase diagram of
the HF metal CeCoIn$_5$ shown in fig. \ref{fig1}. The latter
substance is a $d$ - wave superconductor with $T_c=2.3$ K, while a
field tuned QCP with a critical field of $H_{c0}=5.1$ T coincides
with $H_{c2}$, the upper critical field where superconductivity
vanishes \cite{pag,pag2,bi}. We note that in some cases $H_{c0}=0$.
For example, CeRu$_2$Si$_2$ shows no magnetic ordering down to
lowest temperatures \cite{takah}. Therefore, in our simple HF model
$H_{c0}$ can be treated as a fitting parameter. Under the
application of magnetic field $H_{c0}$, CeCoIn$_5$ demonstrates the
NFL behavior down to $T=0$ \cite{pag1,shag}. It also follows from
the above consideration that $H_{c0}\simeq H_{c2}$ is an accidental
coincidence. Indeed, $H_{c2}$ is determined by $g$ which in turn is
given by the coupling of electrons with magnetic, phonon, etc
excitations rather than by $H_{c0}$. As a result, under application
of a pressure which influences differently $g$ and $H_{c0}$, the
above coincidence will be removed in agreement with facts
\cite{ronn}.
At relatively high temperatures, the superconducting-normal phase
transition in CeCoIn$_5$ shown by the solid line in the right panel
of fig. \ref{fig1} is of the second order \cite{bian,izawa} and $S$
and the other thermodynamic quantities are continuous at the
transition temperature $T_c(H)$. Since $H_{c2}\simeq H_{c0}$, upon
the application of magnetic field, the HF metal transits to its NFL
state down to lowest temperatures as seen from fig. \ref{fig1}. As
long as the phase transition is of the second order, the entropy of
the superconducting phase $S_{\rm SC}(T)$ coincides with the
entropy $S_{\rm NFL}(T)$ of NFL state,
\begin{equation}\label{pq17}
S_{\rm SC}(T\to T_c(H))=S_{\rm NFL}(T\to T_c(H)).
\end{equation}
Since $S_{\rm SC}(T\to 0)\to 0$, eq. (\ref{pq17}) cannot be
satisfied at sufficiently low temperatures due to the presence of
temperature-independent term $S_0$. Thus, in accordance with
experimental results \cite{bian,izawa}, the second order phase
transition converts to first order one below some temperature
$T_{0}(H)$. To estimate $T_{0}(H)$, we use the scaling idea of
Volovik (see ref. \cite{volvol} for details), who derived
interpolation formula for the entropy of a $d$ - wave
superconductor in a magnetic field $H$, while $S_{\rm NFL}$ has
been estimated in \cite{khzvya}. As a result, upon using eq.
(\ref{pq17}), we obtain $T_0(H)/T_c\simeq 0.3$. This point
coincides pretty well with experimental value, shown on the fig.
\ref{fig1}. We note that the prediction that the superconducting
phase transition may change its order had been made in the early
1960-s \cite{maki}. Being based on the thermodynamic
considerations, our proof is robust and can be expanded on cases
when the superconducting phase is replaced by another ordered
state. Namely, if the superconducting phase were replaced by some
other ordered phase separated from the NFL phase by the second
order phase transition at $H=0$, then at some temperature $T_0(H)$
this phase transition should change its order. It follows from
above consideration, that NFL phase has the temperature independent
entropy term $S_0$. Since in the ordered phase the Nernst theorem
($S\to 0$ as $T\to 0$) should hold, we conclude that there is the
entropy step (from $S_0$ to zero) as $T\to 0$ while a system
traverses the phase transition line from ordered phase to the NFL
one. This means that this phase transition should change its order
at $T_0(H)$. For example, we predict that the AFM phase transition
in YbRh$_2$Si$_2$ with $T_N(H)$ (representing the field dependence
of N\'eel temperature) should become first order at $T\leq T_0(H)$,
where $T_{0}(H)$ is some finite temperature. Under constant entropy
(adiabatic) conditions, there should be a temperature step as a
magnetic field crosses the above phase boundary due to the above
thermodynamic inequality. Indeed, the entropy jump would release
the heat, but since $S=const$ the heat is absorbed, causing the
temperature to decrease in order to keep the constant entropy of
the NFL state. Note that the minimal jump is given by the
temperature-independent term $S_0$, which can be quite large so
that the corresponding HF metal can be used as an effective cooler
at low temperatures.

The entropy $S_{\rm NFL}$ determines the anomalous behavior of
CeCoIn$_5$ in the NFL region of the phase diagram. The term
$S_0\sim x(p_f-p_i)/p_F$ can be determined from the experimental
data on spin susceptibility (following Curie law) and the specific
heat jump $\Delta C$ at $T_c$ \cite{khzvya}. In HF metals like
CeCoIn$_5$ the normalized jump $\Delta C/C_n\simeq 4.5$ is
substantially higher than the ordinary BCS value \cite{petr}, where
$C_n$ is the specific heat of a normal state. The specific heat
jump is not proportional to $T_c$ and is related to the fermion
condensate parameter $\delta p_{\rm FC}=(p_f-p_i)/p_F\sim S_0/x$,
therefore the normalized jump $\Delta C/C_n$ can be large
\cite{khzvya,ams1}. This estimation gives $\delta p_{FC}\simeq
0.044$ \cite{khzvya}. The entropy $S_{\rm NFL}$ determines also
both thermal expansion coefficient $\alpha=-\partial S/\partial P$
and Gr\"uneisen ratio $\Gamma=\alpha/C_n$ \cite{zver,alp,khzvya}.
Since the entropy has the temperature independent part $S_0$, the
thermal expansion coefficient $\alpha\simeq-\partial S_0/\partial
P$ becomes temperature independent at low temperatures. Therefore,
at $T \to 0$, $\alpha(T)\to const$, while the specific heat
$C_n(T)\to 0$. As a result, $\Gamma(T \to 0)$ diverges in
coincidence with the facts \cite{pag1}.

Now we consider the LFL behavior tuned by a magnetic field $H\geq
H_{c0}$. The LFL regime is characterized by the temperature
dependence of the resistivity, $\rho(T)=\rho_0+A(H)T^2$, with
$\rho_0$ being the temperature independent part and $A(H)$ is the
scattering coefficient. Since the NFL behavior of CeCoIn$_5$
coincides with that of YbRh$_2$(Si$_{0.95}$Ge$_{0.05}$)$_2$ and
YbRh$_2$Si$_2$ \cite{cust,geg2} we would expect that  the LFL
behavior of these substances also coincide. For example, in
YbRh$_2$(Si$_{0.95}$Ge$_{0.05}$)$_2$ the scattering coefficient
diverges as $A(H)\propto (H-H_{c0})^{-1}$ \cite{cust} while in
CeCoIn$_5$ it diverges as $A(H)\propto (H-H_{c0})^c$ with the
exponent $c\simeq -4/3$ \cite{pag,pag2,bi}. In magnetic fields, the
exponent $c=-1$ characterizes the function $A(H)$ of HF liquid with
FC \cite{shag}, while the exponent $c=-4/3$ describes  the function
$A(H)$ of HF liquid on the disordered side of FCQPT
\cite{shag,clark,shag1}. To understand this striking change in the
behavior of CeCoIn$_5$, we recall that FC has just appeared in this
substance since $\delta p_{FC}=(p_f-p_i)/p_F \simeq 0.044\ll 1$. As
soon as the magnetic field is sufficiently high, $H\geq H_{\rm
cr}$, ($H_{\rm cr}$ is a critical field moving the HF liquid from
the ordered side of FCQPT to the disordered side), Zeeman splitting
$\delta p_F=(p_{F1}-p_{F2})/p_F$ of the two Fermi surfaces of HF
liquid exceeds the condensate parameter, $\delta p_F\geq \delta
p_{FC}$, and the HF liquid with FC becomes LFL placed on the
disordered side near QCP. Here $p_{F1}$ and $p_{F2}$ are the Fermi
momenta of the two Fermi surfaces formed by the application of a
magnetic field. The splitting can be estimated as $p^2_F\delta
p_F/M^*(H)\sim H\mu_B$, where $\mu_B$ is the Bohr magneton. Taking
into account that $A(H)\propto (M^*(H))^2$ we obtain
$(H_{cr}-H_{c0})/H_{c0}\sim (c_1\delta p_F)^{3}$. Our estimations
of the coefficient $c_1$ based on the experimental function $A(H)$
show that $c_1\sim 5$, and we obtain that the reduced field
$(H_{cr}-H_{c0})/H_{c0}\sim (c_1\delta p_F)^3\simeq 0.02$. Thus, we
can safely suggest that the reduced field of $0.02$ is much smaller
than the minimal reduced field $0.1$ where $A(H)$ measurements have
been carried out in ref. \cite{pag,bi}. As a result, the electronic
system of CeCoIn$_5$ is placed on the disordered side of FCQPT by
the application of such a high field and reveals $A(H)\propto
(H-H_{c0})^{-4/3}$. We can see from the left panel of fig.
\ref{fig1} that the coefficient $B(H)$ has the same critical field
dependence. Here $B(H)$ stands for the $T^2$-dependent contribution
to the thermal resistivity and is related to $A(H)$ by a
field-independent factor, $A(H)/B(H)\simeq 0.47$, as it should be
in the case of ordinary metals \cite{pag,pag2} and HF metals
demonstrating the LFL behavior. At sufficiently low temperatures
and decreasing field when $H<H_{cr}$, we predict that CeCoIn$_5$
demonstrates the LFL behavior while the exponent $c$ will change
from $c=-4/3$ to $c=-1$.

At low temperatures and $H\sim H_{cr}$, the system remains in the
LFL regime, but at elevated temperatures there exists a temperature
$T^*(H)$ where the influence of FC related to $S_0$ is recovered
and the NFL behavior is restored. To calculate the function
$T^*(H)$, we note that the effective mass $M^*$ cannot be changed
at $T^*(H)$. Since at $T>T^*(H)$ the effective mass $M^*(T)\propto
1/T$ \cite{noz} and at $T<T^*(H)$, $M^*(H)\propto
(H-H_{c0})^{-2/3}$, we have
\begin{equation}\label{tr1}
T^*(H)\propto (H-H_{c0})^{2/3}.
\end{equation}
The function $T^*(H)$ given by eq. (\ref{tr1}) is represented by
the dotted line in the right panel of fig. \ref{fig1}. In high
magnetic fields, $H\gg H_{cr}$, there is a new crossover line
because the effective mass starts to depend on temperature as
$M^*(T)\propto T^{-2/3}$ \cite{clark,shag1} and $T^*(H)$ becomes
\begin{equation}\label{tr2}
T^*(H)\propto (H-H_{c0}).
\end{equation}
The crossover line given by eq. (\ref{tr2}) is represented by the
solid line in fig. \ref{fig1} (right panel). As it is seen from
fig. \ref{fig1}, the behavior of these lines described by eqs.
(\ref{tr1}) and (\ref{tr2}) are in good agreement with experimental
facts \cite{pag,pag2}. Eventually, when $T_f>T>T^*(H)$ the
influence of FC determined by $S_0$ restores and the system
demonstrates the NFL behavior with $M^*\propto 1/T$ and
$\rho(T)\propto T$ \cite{shag,shag1}. Here $T_f$ is the temperature
at which the influence of FC vanishes. For example, it can be
estimated by using the condition, $S_0\ll S(T_f)$. The NFL behavior
related to the $S_0$ term can also be observed in measurements of
tunneling conductivity and dynamic conductance which are expected
to be noticeably asymmetrical with respect to the change of voltage
bias from $V$ to $-V$ in HF liquid with FC \cite{tun}. Such
asymmetrical conductivity was recently observed experimentally in
CeCoIn$_5$ \cite{park}. The behavior of the conductivity can be
specific when the HF metal transits from its LFL state induced by
the application of magnetic field to NFL one at elevated $T$. We
predict that in the case of CeCoIn$_5$ the conductivity being
symmetrical in the LFL regime becomes gradually asymmetrical
reaching its maximum in the NFL state at elevated temperatures when
$T>T^*(H)$ and eventually vanishes.

In summary, we have presented for the first time theoretical
description of the whole phase diagram of ${\rm CeCoIn_5}$
including the change of the second order superconducting phase
transition to the first one under the application of rising
magnetic field. We have shown that quasiparticles survive down to
lowest temperatures. Our description of the HF metal ${\rm
CeCoIn_5}$ based on the notion of quasiparticles and FCQPT is in
good agreement with facts.

We thank P. Coleman for stimulating discussions. This work was
supported in part by RFBR, project No. 05-02-16085. The visit of
VRS to Clark Atlanta University has been supported by NSF through a
grant to CTSPS.

\end{document}